\documentstyle[epsfig,amsfonts,amssymb,12pt]{article}

\makeatletter
\@addtoreset{equation}{section}

\def\ut#1{\rlap{\lower1ex\hbox{$\sim$}}{#1}}

\def\SL{{\rm SL}}

\newcommand{\R}{{\Bbb R}}

\newcommand{\Z}{{\Bbb Z}}

\begin{document} 
	
\title{A Ponzano-Regge model of Lorentzian 3-Dimensional gravity}
\author {Laurent {\sc Freidel}\thanks{{\tt freidel@ens-lyon.fr}}
$\hspace{2mm}{}^a$
\\[10pt]
${}^a${\it Laboratoire de Physique}  \\
{\it Ecole Normale Sup\'erieure de Lyon} \\
{\it 46, all\'ee d'Italie, 69364 Lyon Cedex 07, France 
\thanks{{URA 5672 du CNRS}}}\\
}
\date{} 
\maketitle

\begin{abstract}
We present the construction of the partition function 
of 3-dimensional gravity in the Lorentzian regime as a state sum 
model over a triangulation. This generalize the work of Ponzano 
and Regge to the case of Lorentzian signature.
\end{abstract}

	\section{3D Gravity}

 It is well known that, in dimension three, gravity is a topological 
theory, no graviton are present  and the number of 
effective degrees of freedom is finite dimensional.
This manifests in the fact that the dynamics  of Lorentzian (resp. Euclidean) 
gravity  can be described using the   topological  
$SO(2,1)$  (resp. $SO(3)$) $BF$ theory (\ref{BF}) whose action is  
given by
\begin{eqnarray}\label{BF}
S[B,A]=\int_{M} {\rm Tr}(B\wedge F(A)),
\end{eqnarray}
where  $B$ is a Lie algebra valued one form, $A$ is a gauge connection 
and $F(A)$  its Lie algebra valued curvature.

For this type of theory it is easy to see  that the unconstrained  phase space is 
given by a pair of `electric and magnetic fields' $(\bar{B},\bar{A})$ 
which are the restriction of the $(B,A)$ fields on a two 
dimensional spatial surface. 
We can therefore take the wave functional as being functional of the 
connection $\psi(\bar{A})$.
The main purpose of this paper is to give a computation of
transition amplitude between choosen initial and final states in the 
lorentzian case.
More precisely, lets consider a 3 dimensional (space-time) 
manifold $M$  with boundary $\partial M$. When $\partial M$ can be 
decomposed into an in and out part 
$\partial M= \partial M_0 \cup \partial M_1$, denoting 
$\bar{A}_0$ (resp. $\bar{A}_1$) the restriction of the connection $A$ on 
$\partial M_0$ (resp. $\partial M_1$) it is possible to define the 
transition amplitude
\begin{equation}\label{HH}
	<\Phi| \Psi>_{M} = 
	\int {\cal D}A {\cal D}e \bar{\Phi}(\bar{A}_0) {\Psi}(\bar{A}_1)
	e^{{i \over h} S[e,A]}
\end{equation}
In the  case where $M= \Sigma \times I$, this amplitude  was computed by 
Witten \cite{Witt1}: 
\begin{equation}\label{Flat}
	<\Phi| \Psi>_{\Sigma}= 
	\int_{{\cal M}_\Sigma}d\mu(A) {\bar \Phi(A)} \Psi(A).
\end{equation}
Here the integral is over the moduli space of flat connections
${\cal M}_\Sigma$ and $d\mu(A)$ is  the symplectic measure on this space.

It turns out that in the case of Euclidean gravity the  computation 
of transition amplitudes (\ref{HH}) was already obtained 
in the sixties by Ponzano and Regge using  recoupling 
coefficients  of $SO(3)$ \cite{PR}.
Let $\Delta$ be a triangulation of the manifold $M$ inducing on the 
boundary a triangulation $\bar{\Delta}_0$ (resp $\bar{\Delta}_1$) of 
$\partial M_0$ (resp. $\partial M_1$).
Lets  label the edges $e$ of the triangulation $\Delta$  by irreducible
representations  of $SO(3)$ with spin $j_{e}\leq k$, where $k$ is a 
regularisation parameter.
In the triangulation $\Delta$ we distinguish two type of edges: the interior 
edges $e$ which do not lie on the boundary and the boundary edges 
$\bar e$ which are edges of the boundary triangulations $\bar{\Delta}_0$ 
or $\bar{\Delta}_1$.
Given a tetrahedra $T$ whose edges are labeled by representations 
$j_e,j_{\bar{e}}$ we denote $T(j_e,j_{\bar{e}})$ the corresponding 
normalized   6-j symbol.
The Ponzano-Regge amplitude associated with the triangulation 
$\Delta$ and the coloring of the boundary triangulation $j_{\bar e}$
is given by:
\begin{equation}
	\label{PR}	
Z(\Delta,  j_{\bar e}) =
\sum_{j_e} \prod_{v} {1 \over \Lambda(k)}
\prod_{e} d({j_e})
\prod_{ T} T(j_e,j_{\bar{e}})
\begin{array}{c}
\end{array}
\end{equation}
where $v$ denotes the vertices of the triangulation, $e$ the interior 
edges, $T$ the tetrahedra, $d({j_e})$ denotes the dimension, 
$\Lambda(k)\sim k^{3}$ is a regularisation  constant and the sum is 
over all possible labeling of the triangulation compatible with the 
labeling of the boundary.
The  Ponzano-Regge model is obtained as the limit 
$k\rightarrow \infty$.
Due to the identities satisfy by the 6-j symbol it turns out that 
the amplitude $Z(\Delta, j_{\bar e})$ does not depend on the choice 
of the triangulation $\Delta$, but only on the boundary data
$\Delta_0,\Delta_1$ and the coloring $ j_{\bar e}$.

The link between Witten and Ponzano-Regge quantization can be 
understood using the notion of spin network.
Given     a triangulation $\bar{\Delta}$ of a 2-dimensional surface 
$\Sigma$  we can construct  a trivalent graph $\Gamma$ dual to it. 
Vertices of this graph correspond to the center of the triangles and 
edges of $\Gamma$ intersect the edges of $\Delta$.
Associated with this graph   we can define a kinematical 
Hilbert space $V_{\Gamma}(\Sigma)$  which is 
 the space of spin networks with support 
$\Gamma$ \cite{AL}.
This means that $V_{\Gamma}$
is the space of $L^2$ functions on $G^E$, where 
$E$ denote the number of edges of $\Gamma$, invariant under the action 
of the gauge group acting at vertices of $\Gamma$.
An orthonormal basis of this space is given by spin network 
functionals $\phi_{\Gamma,j} \in V_{\Gamma}$, were $j$ is a coloring 
of the edges of $\Gamma$ by irreducible representation of
$SO(3)$.
Given a connection $A$ on $\Sigma$, we can construct holonomies 
$g_e(A)$ of the connection $A$ along edges of the graph $\Gamma$.
Taking the value of $\phi_{\Gamma,j}$ on these groups elements 
promotes the spin network functional into a wave function 
$\phi_{\Gamma,j}(\bar{A})$. 

The Ponzano-Regge 
model gives an evaluation of the  wave product (\ref{HH}) of 
such wave functions in term of the amplitude \cite{OO1,Rovelli}:
\begin{equation}\label{KK}
	<\phi_{\Gamma_0,j_0}|\phi_{\Gamma_1,j_1}>_{\Delta} :=
Z(\Delta,  j_{\bar e}).
\end{equation}
where $\Delta$ is a triangulation of $M$ inducing a triangulation 
$\bar{\Delta}_{0,1}$  of $\partial M_{0,1}$  dual to $\Gamma_{0,1}$.

The product (\ref{KK}) coincide, up to a global normalisation,
with the Witten product (\ref{Flat}).
\begin{equation}
<\phi_{\Gamma_0,j_0}|\phi_{\Gamma_1,j_1}>_{\Delta}
= <\phi_{\Gamma_0,j_0}|\phi_{\Gamma_1,j_1}>_{\Sigma},
\end{equation}
showing that  the Ponzano Regge model is in fact equivalent to the Witten 
quantization.
This means, for instance, that when the graphs $\Gamma_{0,1}$ do not wrap 
around a non contractible circle of the two dimensional surface then 
the amplitude is given by the ``evaluation'' of graphs 
\begin{equation}
<\phi_{\Gamma_0,j_0}|\phi_{\Gamma_1,j_1}>_{\Delta}
=ev(\phi_{\Gamma_0,j_0}) ev(\phi_{\Gamma_1,j_1})
\end{equation}
where $ev(\Gamma,j) $ is the evaluation of the spin network,
which is the value of the spin network functional $\phi_{\Gamma,j}$ on 
 the identity group element.

\subsection{Euclidean calculation}

There are several arguments leading to the conclusion that the 
Ponzano-regge model computes transition amplitudes for 3D Euclidean 
gravity \cite{OO1,Rovelli}.
We sketch here the argument of Ooguri \cite{OO1}, when there is no boundary. 
The main idea of this construction is to discretise the measure and 
the action of the partition function \cite{KF}. First, one  choose 
a triangulation $\Delta$ of $M$, then to each edge $e$ of $\Delta$ one associate 
a Lie algebra element 
$B_{e}$ which corresponds to the integral of the $B$ field along the 
edge and and a group element $g_{e}$ which corresponds to the holonomy 
of the connection around the edge. $g_{e}$ is the product of group elements 
$g_{f_{1}}\cdots g_{f_{n}}$ where $f_{i}$ denotes the triangles 
meeting at the edge $e$.
One first integrate over the 
 algebra elements $B_{e}$, each integration  produces a delta 
function imposing the holonomy $g_{e}$ of being the identity.
It can be  shown that, due to the Bianchi identity, a regularisation factor 
denoted $\Lambda$ is needed for each vertex of the triangulation.
In the case of $SU(2)$, the delta function on the group can be 
expressed as a sum over the character of finite dimensional 
representations, by the Plancherel formula:
\begin{equation}
	\delta(g) =\sum_j d_{j} \chi_j(g).
\end{equation}
The sum is over all spins, $d_{j}$    denotes the 
dimension of  the spin $j$ representation and $\chi_j(g)$  denotes the 
trace of a  group element $g$ in this  representation.

The computation of the amplitude therefore reduces to 
\begin{eqnarray} \label{part}
	Z(\Delta) = \prod_{v} \Lambda 
\prod_{f} \int dg_f
\prod_{e } \sum_{j_e} 
d_{j_{e}}\chi_{j_e}(g_{e}),
\end{eqnarray}
where the integral are over the group using the 
normalized Haar measure.

Since a face $f$ of the triangulation possesses 3 edges, each $g_{f}$ appears in three 
characters, so the integration over $g_f$  involves integrals over products of 
three matrix elements. 
It is well known that such  integrals are expressed in term of pairs 
of Clebsh-Gordan coefficients.
Therefore, for each   face of the triangulation we obtain  a contribution involving a 
pair of  Clebsh-Gordan, each Clebsh-Gordan being associated with a 
different tetrahedra.
Since  four  faces are glued in a  tetrahedra we 
get a pairing of four Clebsch-Gordan for each tetrahedra.
In order to understand the pairing it is practical to take the spin 
network notation where  Clebsh-Gordan coefficients correspond to 
trivalent vertices and the pairing is described by the tetrahedral 
graph. The evaluation of this graph is the normalized 
6-j symbol.
When there is a boundary and a spin network functional on the boundary 
there are additional integrations with respect to group elements $h$ 
living on the boundary.
If a face belongs to  the boundary, one of the Clebsh-Gordan is 
involved in a 6-j symbol the other Clebsh-Gordan is used  to 
contract the boundary group elements $h$ into a spin network 
functional.

\section{Lorentzian regime}

The computation we carried out for the Euclidean case can be adapted 
to the Lorentzian case. The main ingredient is to reproduce the 
previous computation when the gauge group is no longer compact but is 
given by the $2+1$ Lorentz group $G=SO(2,1) \sim \SL(2,\R) / 
\Z^{2}$.
We recall first some fact about $\SL(2,\R)$ and its representation 
theory.

We saw that a key ingredient to get (\ref{part}) was the
Plancherel formula.  The irreducible representation appearing in the decomposition  
of the delta function on $\SL(2,\R)$ are all unitary and decompose into three 
different series of representation:\\
i) The  {\it{ principal series}}, $T_{(i\rho - 1/2,\epsilon)}$,\\
$\rho >0 $, $\epsilon= 0,1/2$,
$C_\rho = \rho^{2}+{1\over 4} >{1\over 4}$;\\
ii) The {\it{ holomorphic discrete series}},  $T_l^+ $,\\
$l = 0, {1\over 2}, 1, \cdots$, $C_l=-l(l+1)\leq 0$; \\
iii) the {\it{anti- holomorphic discrete series}}; $T_l^- $, \\
$l = 0, {1\over 2}, 1, \cdots$, 
$C_l=-l(l+1) \leq 0$,\\
where $C$ the Casimir.

Denoting by $\chi_{(\rho,\epsilon)}$, $\chi_{l}^{\pm}$ the 
corresponding characters  
the delta function on the group $\SL(2,\R)$ can be expressed as
\begin{eqnarray}
(\pi)^2 \delta(g) = \nonumber 
\sum_{l} (2l+1) \{ \chi_l^+(g) +\chi_l^-(g) \} + \\
\sum_{\epsilon=0,1/2} \int_{0}^{+\infty}d\rho \chi_{(\rho,\epsilon)}(g) 
  \mu(\rho,\epsilon). \label{Plancherel}
\end{eqnarray}
The sum is over all positive integer and half-integer $l$. 
The Plancherel weight is given by:
\begin{equation}\label{we}
	\mu(\rho,\epsilon) = 2\rho \tanh(\pi \rho +i \epsilon \pi)
\end{equation}

\subsection{Evaluation of the Path integral}

In order to compute the partition function we proceed as in the 
Euclidean case  and we get a expression of the partition functional 
similar to the one presented in (\ref{part}).
There is however two differences. The first one is, of course, that we 
have to replace the sum over the spin $j_e$ by the expansion 
appearing in the Plancherel formula (\ref{Plancherel}). The second 
difference concerns the measure of integration. The 
integrand is a sum of characters and  is  invariant under gauge 
transformation, this means that there is redundant integrals in 
(\ref{part}) that we have to gauge out. This was automatically taken 
into account in the compact case since, in that case, the volume of the  group 
is one. 
The gauge symmetry appearing in (\ref{part}) and the gauge fixing can be describe as follows.
First, consider the lattice  which is the one skeleton of the dual 
triangulation $\Delta^{*}$, the edges of this lattice correspond to the 
faces of the triangulation and the vertices to the tetrahedra.
The gauge theory we are considering is a gauge theory on this lattice. 
As usual, gauge variables  are associated with the edges of the graph and the gauge 
group is acting at the vertices.
In order to gauge fix the action we choose a set of link $T$, called a 
maximal tree,  which does not contain any loop and which is maximal, 
in the sense that it cannot be extend without creating a loop.
Then, we fix the value of group elements on this tree (for instance to 
be one) and integrate the remaining  gauge variables.
This integration can be obtain using the recoupling theory of $\SL(2,\R)$.
The result of the computation can be presented as follows.

For each edge $e$ of the triangulation we choose  an 
orientation of $e$ and  a number  $c_{e}$ belonging to $\{-,0,+\}$.
We consider that a choice of orientation and labeling $c_{e}$ is 
equivalent to the choice of the reverse orientation with the 
labeling $-c_{e}$.   
We call such a labeling of the triangulation a {\it  causal 
structure} $c$ of $\Delta$ and we say that a edge labeled by $0$ is  spacelike, 
an edge labeled by $+$ is future timelike and by $-$ is past timelike.
We obtain that  the partition 
function in the Lorentzian case is given by
\begin{eqnarray}
	Z(\Delta) = \sum_{c} Z(\Delta,c),
\end{eqnarray}
where the sum is over all causal structures $c$ of $\Delta$ and 
\begin{eqnarray}\label{Lor}
Z(\Delta,c) = 
\sum_{l_{e^{+}}} \prod_{e^{+}}  d(l_{e^{+}}) \sum_{l_{e^{-}}} 
\prod_{e_{-}}   d(l_{e^{-}}) \\
\sum_{\epsilon_{0}} \int d\rho_{e_{0}} \prod_{e_{0}} \mu(\rho_{e_{0}},\epsilon_{e_{0}})
\prod_{T} T(l_{e^{-}}, l_{e^{+}} ,\rho_{e_{0}}).
\end{eqnarray}
Where the sum sum over the holomorphic discrete series for  future 
timelike edges, the anti-holomorphic serie for  past timelike  edges and 
 over the principal series for  spacelike  edges.
$d(l_{e^{+}})= 2l_{e^{+}} +1$, $\mu$ is the plancherel weight 
(\ref{we}) and 
$T(l_{e^{-}}, l_{e^{+}} ,\rho_{e^{0}}) $ is the 6-j symbol of $\SL(2,\R)$.
It cannot be defined as in the $SU(2)$ case by the contraction of 
four Clebsh-Gordan coefficients since such contraction would be divergent 
but it is  defined as the matrix element of a unitary 
transformation  on the space of invariant operators acting on 
the tensor product of four unitary representation of $\SL(2,\R)$.
This 6-j symbol can be zero depending on admissibility 
conditions satisfied by the three representations labeling the edges 
of a triangle. 
When the orientation of the triangle is given from the numbering of its 
vertices the admissible triples are \cite{VK}:\\
$(l_{1}^{+}, l_{2}^{+}, l_{3}^{+})$ with  $l_{3} > l_{1}+l_{2}$ 
and $l_{1}+l_{2} +l_{3} \in \Z $, \\
$(l_{1}^{-}, l_{2}^{-}, l_{3}^{-})$ with  $l_{3} > l_{1}+l_{2}$ 
and $l_{1}+l_{2} +l_{3} \in \Z $, \\
$((\rho_{1},\epsilon_{1}), l_{2}^{+}, l_{3}^{-})$ with 
$l_{1} + l_{2} + \epsilon \in \Z $, \\
$((\rho_{1},\epsilon_{1}), (\rho_{2},\epsilon_{2}), l_{3}^{\pm})$ with 
$\epsilon_{1} + \epsilon_{2} + l_{3} \in \Z $, \\
$((\rho_{1},\epsilon_{1}), (\rho_{2},\epsilon_{2}), (\rho_{3},\epsilon_{3})$ with 
$\epsilon_{1} + \epsilon_{2} + \epsilon_{3} \in \Z $.

In the definition of the partition function (\ref{Lor}) a 
regularisation and a limiting procedure is understood exactly as in 
the Euclidean case. We first constrain the sums and integrals to spins 
$l^{\pm}\leq k$, $\rho \leq k$, then add a regularisation factor at the 
vertices of the triangulation before taking the limit.  
The amplitudes $Z(\Delta)$ do not depend on the 
choice of the triangulation. This is not true in general for the 
causal partition function $Z(\Delta,c)$. There is however a important 
exception which arise if the causal structure $c$ is such that all 
edges are labeled $+$ (resp. $-$). In that case the causal partition 
function $Z(\Delta,+)$ (resp. $Z(\Delta,-)$) is invariant under the 
choice of triangulation. 

In Lorentzian geometry a vector in $\R^{3}$ can be spacelike timelike or 
null. This different alternatives shows up in the partition function 
as  the possibility to choose which type of representation we assign to 
the edges. The fact that we assign principal series to spacelike edges 
and discrete series for timelike edges can be justified by several 
arguments which we can just outline here.
First, with this geometrical understanding of representation, the 
admissibility rules are consistent with the rule of vector addition.
Second, it is  consistent with the Kirillov correspondence between 
representation of $SL(2,\R)$ and orbits in $R^{3}$ since in this 
correspondence the discrete series are associated with two-sheeted 
hyperboloid which are orbit of a timelike vector.

\end{document}